\documentclass{article}
\usepackage{graphicx}
\graphicspath{%
    {converted_graphics/}% inserted by PCTeX
    {C:/Dropbox/1-kgl-top/Papers/2014/Interferometer/}% inserted by PCTeX
}
\begin{document}

\begin{center}
{\huge A Laser Interferometer}\vskip8pt

{\huge for the Undergraduate Teaching Laboratory}\vskip4pt

{\huge Demonstrating Picometer Sensitivity}\vskip12pt

{\Large Kenneth G. Libbrecht and Eric D. Black}\vskip4pt

{\large Department of Physics, California Institute of Technology}\vskip-1pt

{\large Pasadena, California 91125}\vskip-1pt

\vskip18pt

\hrule\vskip1pt \hrule\vskip14pt
\end{center}

\textbf{Abstract.} We describe a laser interferometer experiment for the
undergraduate teaching laboratory that achieves picometer sensitivity in a
hands-on table-top instrument. In addition to providing an introduction to
interferometer physics and optical hardware, the experiment also focuses on
precision measurement techniques including servo control, signal modulation,
phase-sensitive detection, and different types of signal averaging. After
students assemble, align, and characterize the interferometer, they then use
it to measure nanoscale motions of a simple harmonic oscillator system, as a
substantive example of how laser interferometry can be used as an effective
tool in experimental science.

\section{Introduction}

Optical interferometry is a well-known experimental technique for making
precision displacement measurements, with examples ranging from Michelson
and Morley's famous aether-drift experiment to the extraordinary sensitivity
of modern gravitational-wave detectors. With careful attention to various
fundamental and technical noise sources, displacement sensitivities of
better than $10^{-19}$ m/Hz$^{-1/2}$ have been demonstrated (roughly $%
10^{-13}$ wavelengths of light) \cite{LIGO, TNI1, TNI2}, and additional
improvements are expected in the near future.

In the undergraduate laboratory, however, the use of optical interferometry
as a teaching tool has not kept pace with the development of modern
measurement techniques. Interferometer experiments in the teaching lab often
stop with the demonstration of visible fringes followed by displacement
measurement using basic fringe counting. We believe there is a substantial
pedagogical opportunity in this area to explore modern precision measurement
concepts using an instrument that is visual and tactile, relatively easy to
understand, and generally fun to work with.

While there are numerous examples of precision laser interferometry in the
literature \cite{IFO0, IFO1, IFO2, IFO3, IFO4}, these instruments are
somewhat complex in their design and are therefore not optimally suited for
a teaching environment. Below we describe a laser interferometer designed to
demonstrate precision physical measurement techniques in a compact apparatus
with a relatively simple optical layout. Students place some of the optical
components and align the interferometer, thus gaining hands-on experience
with optical and laser hardware. The alignment is straightforward but not
trivial, and the various interferometer signals are directly observable on
the oscilloscope. Some features of the instrument include:\ 1) piezoelectric
control of one mirror's position, allowing precise control of the
interferometer signal; 2) the ability to lock the interferometer at its most
sensitive point; 3) the ability to modulate the mirror position while the
interferometer is locked, thus providing a displacement signal of variable
magnitude and frequency; 4) phase-sensitive detection of the modulated
displacement signal, both using the digital oscilloscope and using basic
analog signal processing.

In working with this experiment, students are guided from micron-scale
measurement precision using direct fringe counting to picometer precision
using a modulated signal and phase-sensitive signal averaging. The end
result is the ability to see displacement modulations below one picometer in
a 10-cm-long interferometer arm, which is like measuring the distance from
New York to Los Angeles with a sensitivity better than the width of a human
hair!

Once the interferometer performance has been explored, students then
incorporate a magnetically driven oscillating mirror in the optical layout.
Observation and analysis of nanometer-scale motions of the high-Q oscillator
reveal several aspects of its behavior, including: 1) the
near-resonant-frequency response of the oscillator; 2) mass-dependent
frequency shifts; 3) changes in the mechanical Q as damping is added; and 4)
the excitation of the oscillator via higher harmonics using a square-wave
drive signal.

With this apparatus, students learn about optical hardware and lasers,
optical alignment, laser interferometry, piezoelectric transducers,
photodetection, electronic signal processing, signal modulation to avoid
low-frequency noise, signal averaging, and phase-sensitive detection.
Achieving a displacement sensitivity of 1/100th of an atom with a table-top
instrument provides an impressive demonstration of the power of
interferometric measurement and signal-averaging techniques. Further
quantifying the behavior of a mechanical oscillator executing nanoscale
motions shows the effectiveness of laser interferometry as a measurement
tool in experimental science.

\begin{figure}[htb] % float placement: (h)ere, page (t)op, page (b)ottom, other (p)age
  \centering
  % file name: C:/Dropbox/1-kgl-top/Papers/2014/Interferometer/IFOSetup.jpg
  \includegraphics[width=5.0in, keepaspectratio]{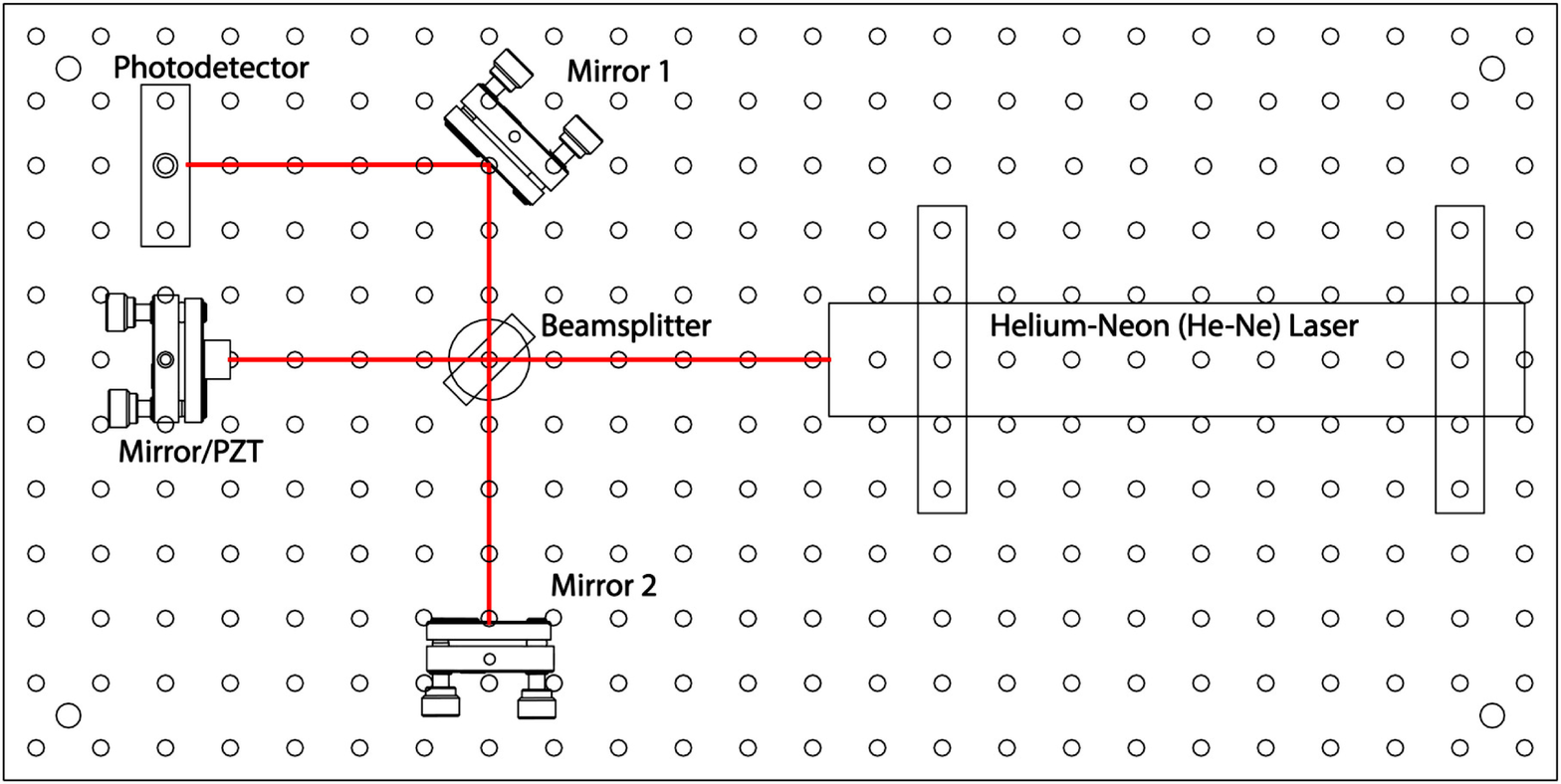}
  \caption{The interferometer optical
layout on an aluminum breadboard with mounting holes on a 25.4-mm grid. The
Mirror/PZT consists of a small mirror glued to a piezoelectric stack mounted
to a standard optical mirror mount. Mirrors 1 and 2 are basic steering
mirrors, and the Beamsplitter is a wedge with a 50:50 dielectric coating.}
  \label{ifolayout}
\end{figure}

\section{Interferometer Design and Performance}

Figure \ref{ifolayout} shows the overall optical layout of the constructed
interferometer. The 12.7-mm-thick aluminum breadboard (Thorlabs MB1224) is
mounted atop a custom-made steel electronics chassis using pliable rubber
vibration dampers, and the chassis itself rests on pliable rubber feet. We
have found that this two-stage seismic isolation system is adequate for
reducing noise in the interferometer signal arising from benchtop
vibrations, as long as the benchtop is not bumped or otherwise unnecessarily
perturbed.

The Helium-Neon laser (Meredith HNS-2P) produces a 2mW linearly polarized
(500:1 polarization ratio) 633-nm beam with a diameter of approximately 0.8
mm, and it is mounted in a pair of custom fixed acrylic holders. The
Beamsplitter (Thorlabs BSW10) is a 1-inch-diameter wedged plate beamsplitter
with a broadband dielectric coating giving roughly equal transmitted and
reflected beams. It is mounted in a fixed optical mount (Thorlabs FMP1)
connected to a pedestal post (Thorlabs RS1.5P8E) fastened to the breadboard
using a clamping fork (Thorlabs CF125). Mirrors 1 and 2 (both Thorlabs
BB1-E02) are mounted in standard optical mounts (Thorlabs KM100) on the same
pedestal posts. Using these stout steel pedestal posts is important for
reducing unwanted motions of the optical elements.

The Mirror/PZT consists of a small mirror (12.5-mm diameter, 2-mm thick,
Edmund Optics 83-483, with an enhanced aluminum reflective coating) glued to
one end of a piezoelectric stack transducer (PZT) (Steminc SMPAK155510D10),
with the other end glued to an acrylic disk in a mirror mount. An acrylic
tube surrounds the Mirror/PZT assembly for protection, but the mirror only
contacts the PZT stack. The surface quality of the small mirror is
relatively poor (2-3 waves over one cm) compared with the other mirrors, but
we found it is adequate for this task, and the small mass of the mirror
helps push mechanical resonances of the Mirror/PZT assembly to frequencies
above 700 Hz.

The photodetector includes a Si photodiode (Thorlabs FDS100) with a 3.6mm x
3.6mm active area, held a custom acrylic fixed mount. The custom photodiode
amplifier consists of a pair of operational amplifiers (TL072) that provide
double-pole low-pass filtering of the photodiode signal with a 10-$\mu $sec
time constant. The overall amplifier gain is fixed, giving approximately an
8-volt output signal with the full laser intensity incident on the
photodiode's active area.

The optical layout shown in Figure \ref{ifolayout} was designed to provide
enough degrees of freedom to fully align the interferometer, but no more.
The Mirror/PZT pointing determines the degree to which the beam is
misaligned from retroreflecting back into the laser (described below), the
Mirror 2 pointing allows for alignment of the recombining beams, and the
Mirror 1 pointing is used to center the beam on the photodiode. In addition
to reducing the cost of the interferometer and its maintenance, using a
small number of optical elements also reduces the complexity of the set-up,
improving its function as a teaching tool.

Three of the optical elements (Mirror 1, Mirror 2, and the Beamsplitter) can
be repositioned on the breadboard or removed. The other three elements (the
laser, photodiode, and the Mirror/PZT) are fixed on the breadboard, the only
available adjustment being the pointing of the Mirror/PZT. The latter three
elements all need electrical connections, and for these the wiring is sent
down through existing holes in the breadboard and into the electronics
chassis below. The use of fixed wiring (with essentially no accessible
cabling) allows for an especially robust construction that simplifies the
operation and maintenance of the interferometer. At the same time, the three
free elements present students with a realistic experience placing and
aligning laser optics.

Before setting up the interferometer as in Figure \ref{ifolayout}, there are
a number of smaller exercises students can do with this instrument. The
Gaussian laser beam profile can be observed, as well as the divergence of
the laser beam. Using a concave lens (Thorlabs LD1464-A, f = -50 mm)
increases the beam divergence and allows a better look at the beam profile.
Laser speckle can also be observed, as well as diffraction from small bits
of dirt on the optics. Ghost laser beams from the antireflection-coated side
of the beamsplitter are clearly visible, as the wedge in the glass sends
these beams out at different directions from the main beams. Rotating the
beamsplitter 180 degrees results in a different set of ghost beams, and it
is instructive to explain these with a sketch of the two reflecting surfaces
and the resulting intensities of multiply reflected beams.

\begin{figure}[htb] % float placement: (h)ere, page (t)op, page (b)ottom, other (p)age
  \centering
  % file name: C:/Dropbox/1-kgl-top/Papers/2014/Interferometer/IFOdiagram2.jpg
  \includegraphics[width=3.0in, keepaspectratio]{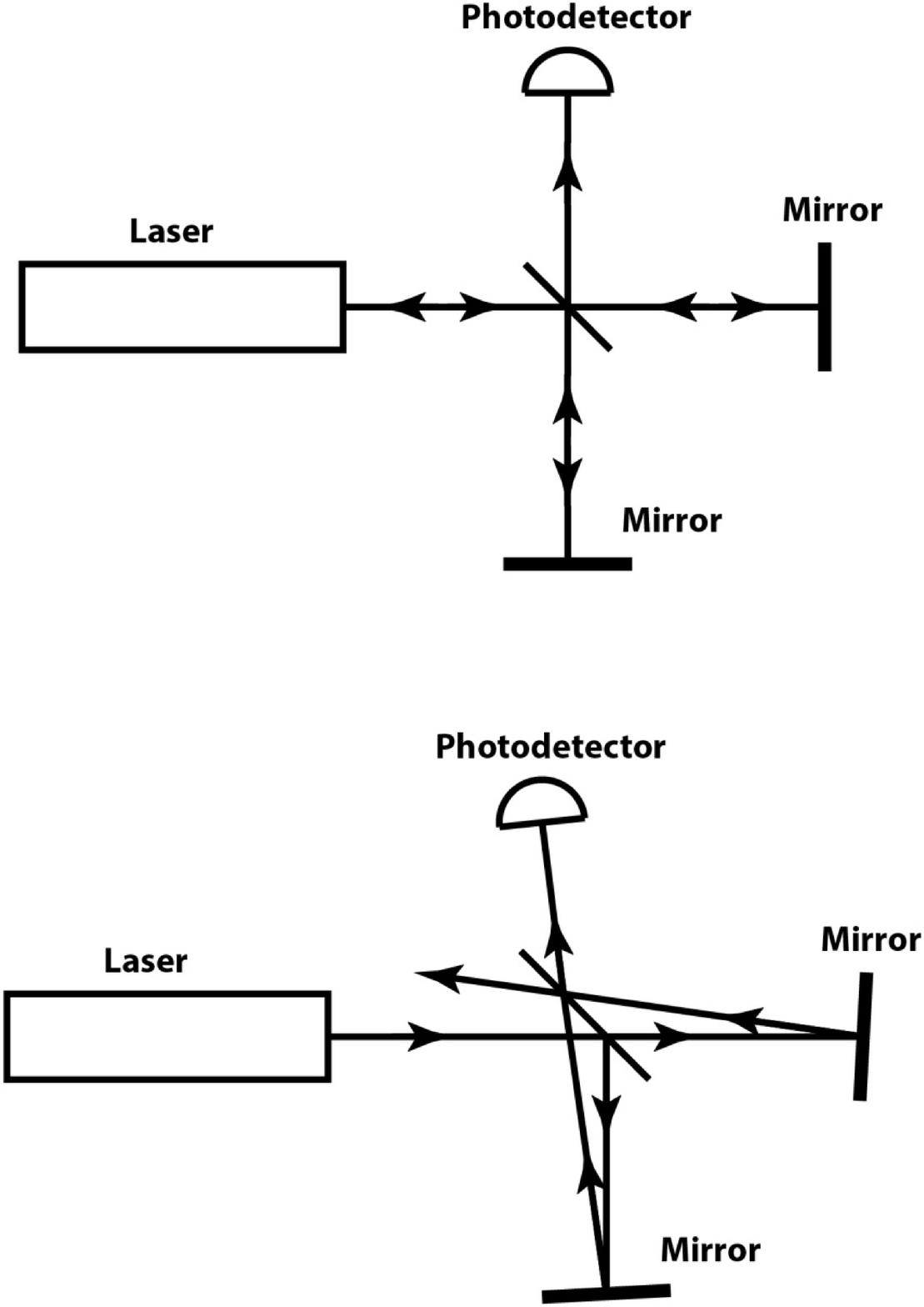}
  \caption{Although the top diagram is
often used to depict a basic Michelson interferometer, in reality this
configuration is impractical. Reflections from the front mirror of the laser
produce multiple interfering interferometers that greatly complicate the
signal seen at the photodetector. In contrast, the lower diagram shows how a
slight misalignment (exaggerated in the diagram) eliminates these unwanted
reflections without the need for additional optical elements. In the
misaligned case, however, complete overlap of the recombined beams is only
possible if the arm lengths of the interferometer are equal.}
  \label{ifoalignment}
\end{figure}

\subsection{Interferometer Alignment}

A satisfactory alignment of the interferometer is straightforward and easy
to achieve, but doing so requires an understanding of how real-world optics
can differ from the idealized case that is often presented. As shown in
Figure \ref{ifoalignment}, retroreflecting the laser beams at the ends of
the interferometer arms yields a recombined beam that is sent directly back
toward the laser. This beam typically reflects off the front mirror of the
laser and reenters the interferometer, yielding an optical cacophony of
multiple reflections and unwanted interference effects. Inserting an optical
isolator in the original laser beam would solve this problem, but this is an
especially expensive optical element that is best avoided in the teaching
lab.

The preferred solution to this problem is to misalign the arm mirrors
slightly, as shown in Figure \ref{ifoalignment}. With our components and the
optical layout shown in Figure \ref{ifolayout}, misaligning the Mirror/PZT
by 4.3 mrad is sufficient that the initial reflection from the Mirror/PZT
avoids striking the front mirror of the laser altogether, thus eliminating
unwanted reflections. This misalignment puts a constraint on the lengths of
the two arms, however, as can be seen from the second diagram in Figure \ref%
{ifoalignment}. If the two arm lengths are identical (as in the diagram),
then identical misalignments of both arm mirrors can yield (in principle)
perfectly recombined beams that are overlapping and collinear beyond the
beamsplitter. If the arm lengths are not identical, however, then perfect
recombination is no longer possible.

The arm length asymmetry constraint can be quantified by measuring the
fringe contrast seen by the detector. If the position $x$ of the Mirror/PZT
is varied over small distances, then the detector voltage can be written%
\begin{equation}
V_{\det }=V_{\min }+\frac{1}{2}(V_{\max }-V_{\min })[1+\cos (2kx)]
\label{detectorvoltage}
\end{equation}%
where $V_{\min }$ and $V_{\max }$ are the minimum and maximum voltages,
respectively, and $k=2\pi /\lambda $ is the wavenumber of the laser. This
signal is easily observed by sending a triangle wave to the PZT, thus
translating the mirror back and forth, while $V_{\det }$ is observed on the
oscilloscope. We define the interferometer fringe contrast to be%
\[
F_{C}=\frac{V_{\max }-V_{\min }}{V_{\max }+V_{\min }} 
\]%
and a high fringe contrast with $F_{C}\approx 1$ is desirable for obtaining
the best interferometer sensitivity.

With this background, the interferometer alignment consists of several
steps: 1) Place the beamsplitter so the reflected beam is at a 90-degree
angle from the original laser beam. The beamsplitter coating is designed for
a 90-degree reflection angle, plus it is generally good practice to keep the
beams on a simple rectangular grid as much as possible; 2) With Mirror 2
blocked, adjust the Mirror/PZT pointing so the reflected beam just misses
the front mirror of the laser. This is easily done by observing any multiple
reflections at the photodiode using a white card; 3) Adjust the Mirror 1
pointing so the beam is centered on the photodiode; 4) Unblock Mirror 2 and
adjust its pointing to produce a single recombined beam at the photodiode;
5) Send a triangle wave signal to the PZT, observe $V_{\det }$ with the
oscilloscope, and adjust the Mirror 2 pointing further to obtain a maximum
fringe contrast $F_{C,\max }.$

\begin{figure}[htb] % float placement: (h)ere, page (t)op, page (b)ottom, other (p)age
  \centering
  % file name: C:/Dropbox/1-kgl-top/Papers/2014/Interferometer/ContrastPlot2.jpg
  \includegraphics[width=3.5in,keepaspectratio]{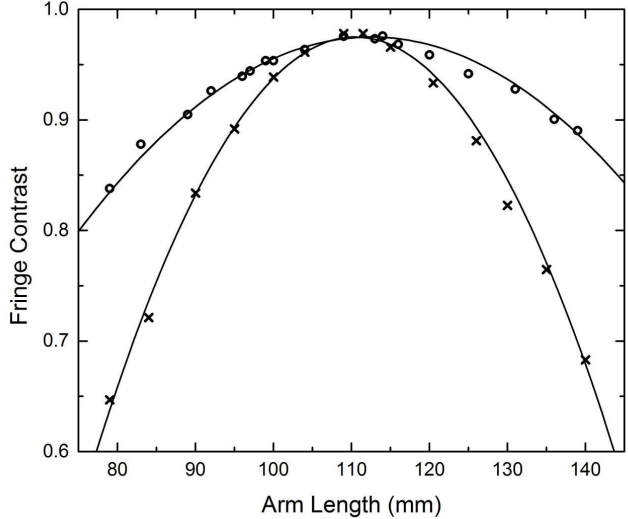}
  \caption{The measured fringe contrast $%
F_{C,\max }$ as a function of the length of the Mirror 2 arm of the
interferometer. For each data point, Mirror 2 was repositioned and
reclamped, and then the Mirror 2 pointing was adjusted to obtain the maximum
possible fringe contrast. The \textquotedblleft X\textquotedblright\ points
were taken with a Mirror/PZT misalignment of 4.3 mrad (relative to
retroreflection), while the circles were taken with a misalignment of 1.3
mrad. The lines show parabolic fits to the data.}
  \label{contrastplot}
\end{figure}

Figure \ref{contrastplot} shows our measurements of $F_{C,\max }$ as a
function of the Mirror 2 arm length when the Mirror/PZT misalignment was set
to 4.3 mrad and the Mirror/PZT arm length was 110 mm. As expected, the
highest $F_{C,\max }$ was achieved when the arm lengths were equal. With
unequal arm lengths, perfect recombination of the beams is not possible, and
we see that $F_{C,\max }$ drops off quadratically with increasing asymmetry
in the arm lengths.

As another alignment test, we misaligned the Mirror/PZT by 1.3 mrad and
otherwise followed the same alignment procedure described above, giving the
other set of data points shown in Figure \ref{contrastplot}. With this
smaller misalignment, there were multiple unwanted reflections from the
front mirror of the laser, but these extra beams were displaced just enough
to miss the active area of the photodetector. In this case we see a weaker
quadratic dependence of $F_{C,\max }$ on the Mirror 2 position, and about
the same $F_{C,\max }$ when the arm lengths are identical.

We did not examine why $F_{C,\max }$ is below unity for identical arm
lengths, but this is likely caused by the beamsplitter producing unequal
beam intensities, and perhaps by other optical imperfections in our system.
The peak value of about 97\% shows little dependence on polarization angle,
as observed by rotating the laser tube in its mount. Extrapolating the data
in Figure \ref{contrastplot} to zero misalignment suggests that the laser
has an intrinsic coherence length of roughly 15 cm. We did not investigate
the origin of this coherence length, although it appears likely that it
arises in part from the excitation of more than one longitudinal mode in the
laser cavity.

The smaller 1.3-mrad misalignment produces a higher fringe contrast for
unequal arm lengths, but this also requires that students deal with what can
be a confusing array of unwanted reflections. When setting up the
interferometer configuration shown in Figure \ref{ifolayout}, we typically
have students use the larger misalignment of 4.3 mrad, which is set up by
observing and then quickly eliminating the unwanted reflections off the
front mirror of the laser. We then ask students to match the interferometer
arm lengths to an accuracy of a few millimeters, as this can be done quite
easily from direct visual measurement using a plastic ruler.

Once the interferometer is roughly aligned (with the 4.3 mrad misalignment),
it is also instructive to view the optical fringes by eye using a white
card. Placing a negative lens in front of the beamsplitter yields a
bull's-eye pattern of fringes at the photodetector, and this pattern changes
as the Mirror 2 pointing is adjusted. Placing the same lens after the
beamsplitter gives a linear pattern of fringes, and the imperfect best
fringe contrast can be easily seen by attempting (unsuccessfully) to produce
a perfectly dark fringe on the card.

\begin{figure}[htb] % float placement: (h)ere, page (t)op, page (b)ottom, other (p)age
  \centering
  % file name: C:/Dropbox/1-kgl-top/Papers/2014/Interferometer/servo.jpg
  \includegraphics[width=4in,keepaspectratio]{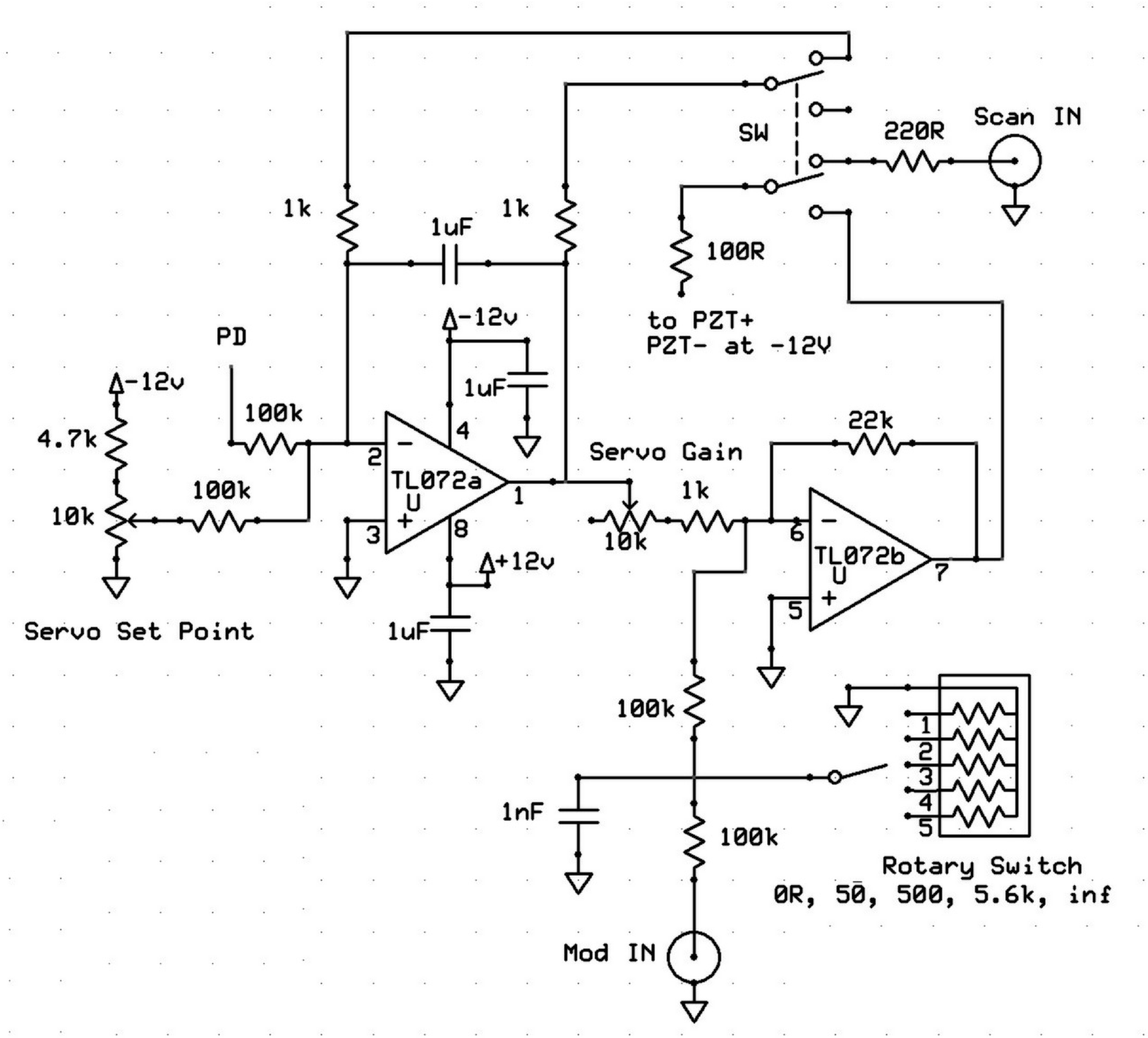}
  \caption{The electronics used to scan,
lock, and modulate the interferometer signal. With switch SW in the SCAN
position, a signal input to the Scan IN port is sent essentially directly to
the PZT. With the switch in the LOCK position, a feedback loop locks the
Mirror/PZT so the average photodiode signal (PD) equals the Servo Set Point.
With the interferometer locked, a signal sent to the Mod IN port
additionally modulates the mirror position. A resistor divider is used to
turn off the modulation or reduce it's amplitude by a factor of 1, 10, 100,
or 1000.}
  \label{servo}
\end{figure}

\subsection{Interferometer Locking}

The interferometer is locked using the electronic servo circuit shown in
Figure \ref{servo}. In short, the photodiode signal $V_{\det }$ is fed back
to the PZT via this circuit to keep the signal at some constant average
value, thus keeping the arm length difference constant to typically much
better than $\lambda /2.$ The total range of the PZT is only about 1 $\mu $m
(with an applied voltage ranging from 0 to 24 volts), but this is sufficient
to keep the interferometer locked for hours at a time provided the system is
stable and undisturbed. Typically the set point is adjusted so the
interferometer is locked at $V_{\det }=(V_{\min }+V_{\max })/2,$ which is
the point where the interferometer sensitivity $dV_{\det }/dx$ is highest.

Note that the detector signal $V_{\det }$ is easily calibrated by measuring $%
\Delta V=V_{\max }-V_{\min }$ on the oscilloscope and using Equation \ref%
{detectorvoltage}, giving the conveniently simple approximation%
\[
\left( \frac{dV_{\det }}{dx}\right) _{\max }\approx \frac{\Delta V}{100 \rm{
nm}} 
\]%
which is accurate to better than one percent. Simultaneously measuring $%
V_{\det }$ and the voltage $V_{PZT}$ sent to the PZT via the Scan IN port
(see Figure \ref{servo}) quickly gives the absolute PZT response function $%
dx/dV_{PZT}$.

The PZT can also be modulated with the servo locked using the circuit in
Figure \ref{servo} along with an external modulation signal. Figure \ref%
{servoon} shows the interferometer response as a function of modulation
frequency in this case, for a fixed input modulation signal amplitude. To
produce these data we locked the interferometer at $V_{\det }=(V_{\min
}+V_{\max })/2$ and provided a constant-amplitude sine-wave signal to the
modulation input port shown in Figure \ref{servo}. The resulting sine-wave
response of $V_{\det }$ was then measured using a digital oscilloscope for
different values of the modulation frequency, with the servo gain at its
minimum and maximum settings (see Figure \ref{servo}).

A straightforward analysis of the servo circuit predicts that the
interferometer response should be given by%
\[
\left\vert \delta V_{det}\right\vert =AG_{1}V_{mod}\left[ 1+\frac{%
AG_{2}}{2\pi \tau \nu }\right] ^{-1/2} 
\]%
where $A(\nu )=dV_{\det }/dV_{PZT}$ includes the frequency-dependent PZT
response, $\nu $ is the modulation frequency, $V_{mod}$ is the
modulation voltage, and the remaining parameters $(G_{1}=0.11;$ $G_{2}=22$
(high gain), $2$ (low gain); $\tau =RC=0.1$ seconds) can be derived from the
servo circuit elements shown in Figure \ref{servo}. Direct measurements
yielded $A(\nu )\approx 3.15,$ where this number was nearly
frequency-independent below 600 Hz and dropped off substantially above 1
kHz. In addition, a number of mechanical resonances in the Mirror/PZT
housing were also seen above 700 Hz. The theory curves shown in Figure \ref%
{servoon} assume a frequency-independent $A(\nu )$ for simplicity.

From these data we see that at low frequencies the servo compensates for the
modulation input, reducing the interferometer response, and the reduction is
larger when the servo gain is higher. This behavior is well described by the
servo circuit theory. At frequencies above about 700 Hz, the data begin to
deviate substantially from the simple theory. The theory curves in principle
contain no adjustable parameters, but we found that the data were better
matched by including an overall multiplicative factor of 0.94 in the theory.
This six-percent discrepancy was consistent with the overall uncertainties
in the various circuit parameters.

\begin{figure}[htb] % float placement: (h)ere, page (t)op, page (b)ottom, other (p)age
  \centering
  % file name: C:/Dropbox/1-kgl-top/Papers/2014/Interferometer/ServoOn2.jpg
  \includegraphics[width=3.5in, keepaspectratio]{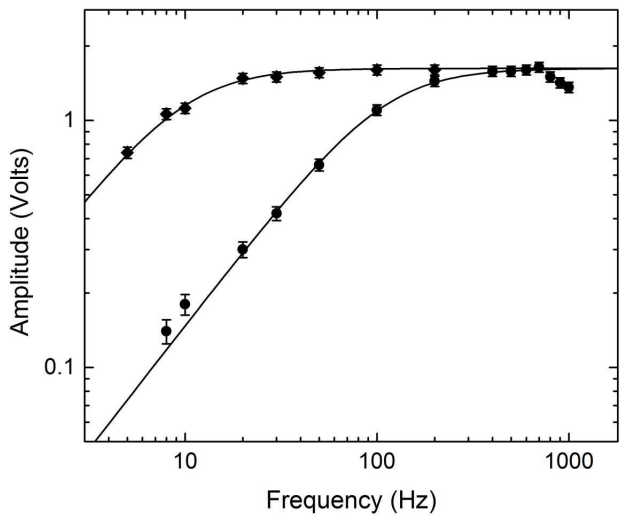}
  \caption{Measurements of the
interferometer response as a function of the PZT modulation frequency, with
the servo locked. The upper and lower data points were obtained with the
servo gain at its lowest and highest settings, respectively, using the servo
control circuit shown in Figure \protect\ref{servo}. The theory curves were
derived from an analysis of the servo control circuit, using parameters that
were measured or derived from circuit elements. To better match the data,
the two theory curves each include an additional multiplicative factor of
0.94, consistent with the estimated overall uncertainty in determining the
circuit parameters.}
  \label{servoon}
\end{figure}

\begin{figure}[htb] % float placement: (h)ere, page (t)op, page (b)ottom, other (p)age
  \centering
  % file name: C:/Dropbox/1-kgl-top/Papers/2014/Interferometer/PSD.jpg
  \includegraphics[width=3.5in, keepaspectratio]{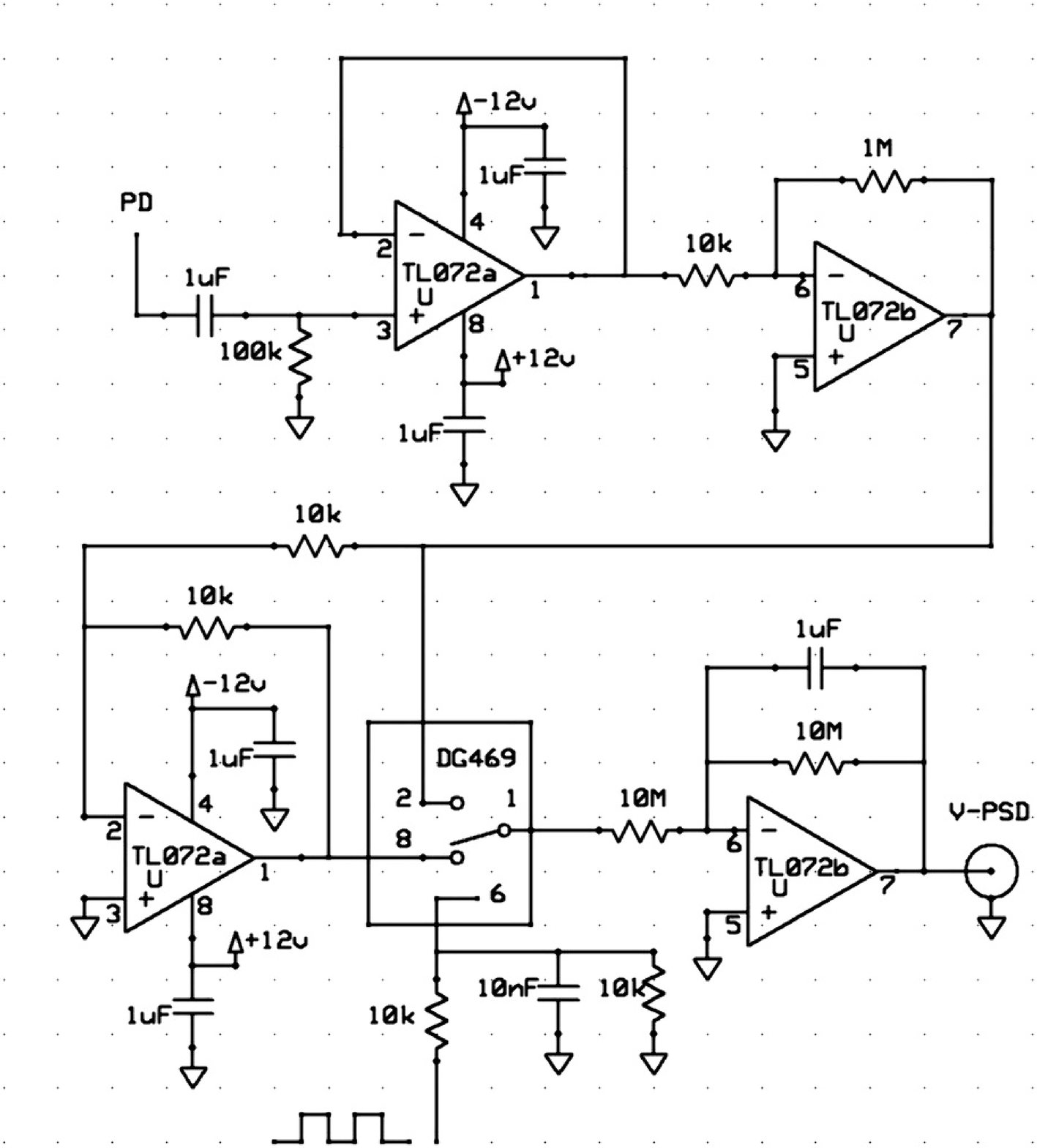}
  \caption{The electronics used to perform
a phase-sensitive detection and averaging of the modulated interferometer
signal. The input signal from the photodiode amplifier (PD) is first
low-pass filtered and further amplified, plus a negative copy is produced
with a $G=-1$ amplifier. An analog electronic switch chops between these two
signals, driven synchronously with the modulation input, and the result is
amplified and averaged using a low-pass filter with a time constant of 10
seconds.}
  \label{psdcircuit}
\end{figure}

\subsection{Phase-Sensitive Detection}

Since the purpose of building an interferometer is typically to measure
small displacement signals, we sought to produce the highest displacement
sensitivity we could easily build in a compact teaching instrument. With the
interferometer locked at its most sensitive point, direct observations of
fluctuations in $V_{\det }$ indicate an ambient displacement noise of
roughly 1 nm RMS over short timescales at the maximum servo gain, and about
4 nm at the minimum servo gain. Long-term drifts are compensated for by the
servo, and these drifts were not investigated further. The short-term noise
is mainly caused by local seismic and acoustic noise. Tapping on the table
or talking around the interferometer clearly increases these noise sources.

To quantify the interferometer sensitivity, we modulated the PZT with a
square wave signal at various amplitudes and frequencies, and we observed
the resulting changes in $V_{\det }.$ The environmental noise sources were
greater at lower frequencies, so we found it optimal to modulate the PZT at
around 600 Hz. This frequency was above much of the environmental noise and
above where the signal was reduced by the servo, but below the mechanical
resonances in the PZT housing.

With a large modulation amplitude, one can observe and measure the response
in $V_{\det }$ directly on the oscilloscope, as the signal/noise ratio is
high for a single modulation cycle. At lower amplitudes, the signal is
better observed by averaging traces using the digital oscilloscope, while
triggering with the synchronous modulation input signal. By averaging 128
traces, for example, one can see signals that are about ten times lower than
is possible without averaging, as expected.

To carry this process further, we constructed the basic phase-sensitive
detector circuit shown in Figure \ref{psdcircuit}, which is essentially a
simple (and inexpensive) alternative to using a lock-in amplifier. By
integrating for ten seconds, this circuit averages the modulation signal
over about 6000 cycles, thus providing nearly another order-of-magnitude
improvement over signal averaging using the oscilloscope. The output $%
V_{PSD} $ from this averaging circuit also provides a convenient voltage
proportional to the interferometer modulation signal that can be used for
additional data analysis. For example, observing the distribution of
fluctuations in $V_{PSD}$ over timescales of minutes to hours gives a
measure of the uncertainty in the displacement measurement being made by the
interferometer.

Our pedagogical goal in including these measurement strategies is to
introduce students to some of the fundamentals of modern signal analysis.
Observing the interferometer signal directly on the oscilloscope is the most
basic measurement technique, but it is also the least sensitive, as the
direct signal is strongly affected by environmental noise. A substantial
first improvement is obtained by modulating the signal at higher
frequencies, thus avoiding the low-frequency noise components. Simple signal
averaging using the digital oscilloscope further increases the signal/noise
ratio, demonstrating a simple form of phase-sensitive detection and
averaging, using the strong modulation input signal to trigger the
oscilloscope. Additional averaging using the circuit in Figure \ref{servo}
yields an expected additional improvement in sensitivity. Seeing the gains
in sensitivity at each stage in the experiment introduces students to the
concepts of signal modulation, phase-sensitive detection, and signal
averaging, driving home the $\sqrt{N}$ averaging rule.

\subsection{Interferometer Response}

\begin{figure}[htb] % float placement: (h)ere, page (t)op, page (b)ottom, other (p)age
  \centering
  % file name: C:/Dropbox/1-kgl-top/Papers/2014/Interferometer/Displacement.jpg
  \includegraphics[width=5in, keepaspectratio]{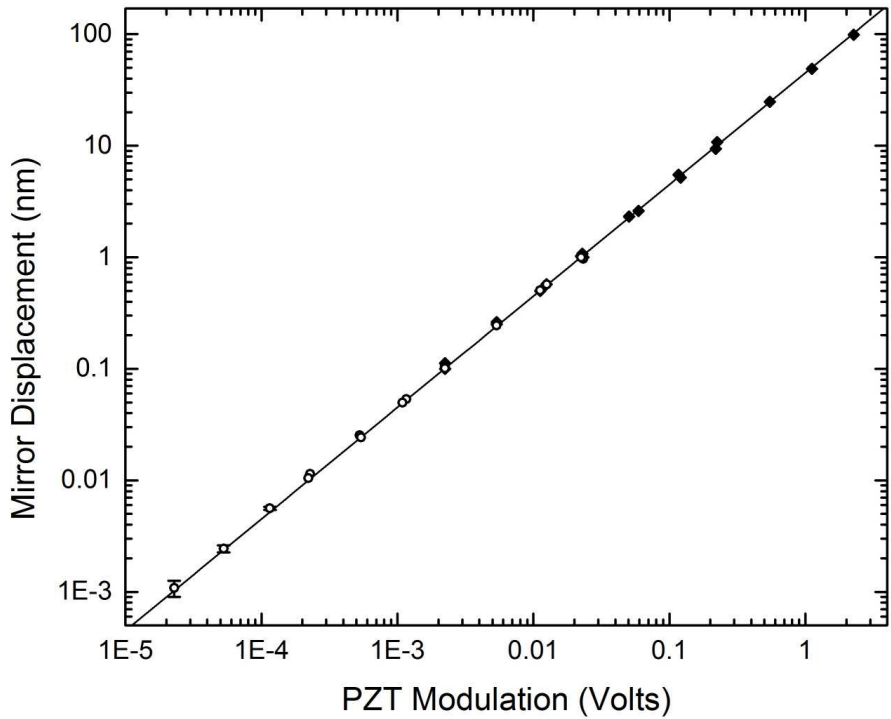}
  \caption{The measured mirror displacement
when the piezoelectric transducer was driven with a square wave modulation
at 600 Hz, as a function of the modulation amplitude. The high-amplitude
points (closed diamonds) were measured by observing the photodiode signal
directly on the oscilloscope, while the low-amplitude points (open circles)
were measured using the phase-sensitive averaging circuit shown in Figure 
\protect\ref{psdcircuit}. The fit line gives a PZT response of 45 nm/volt.
These data indicate that the combined PZT and photodiode responses are quite
linear over a range of five orders of magnitude in amplitude. At the lowest
modulation amplitudes, the noise in the averaged interferometer signal was
below one picometer for 10-second averaging times.}
  \label{displacement}
\end{figure}

Figure \ref{displacement} shows the measured interferometer response at 600
Hz as a function of the PZT modulation amplitude. When the displacement
amplitude was above 0.1 nm, the modulation signal was strong enough to be
measured using the digital oscilloscope's measure feature while averaging
traces. At low displacement amplitudes, the signal became essentially
unmeasurable using the oscilloscope alone, but still appeared with high
signal-to-noise using the $V_{PSD}$ output. The overlap between these two
methods was used to determine a scaling factor between them. The absolute
measurement accuracy was about 5\% for these data, while the 1$\sigma $
displacement sensitivity at the lowest amplitudes was below 1 picometer.
These data indicate that systematic nonlinearities in the photodiode and the
PZT stack response were together below 10 percent over a range of five
orders of magnitude.

\section{Measuring a Simple Harmonic Oscillator}

\begin{figure}[htb] % float placement: (h)ere, page (t)op, page (b)ottom, other (p)age
  \centering
  % file name: C:/Dropbox/1-kgl-top/Papers/2014/Interferometer/WithOscillator.jpg
  \includegraphics[width=5.0in, keepaspectratio]{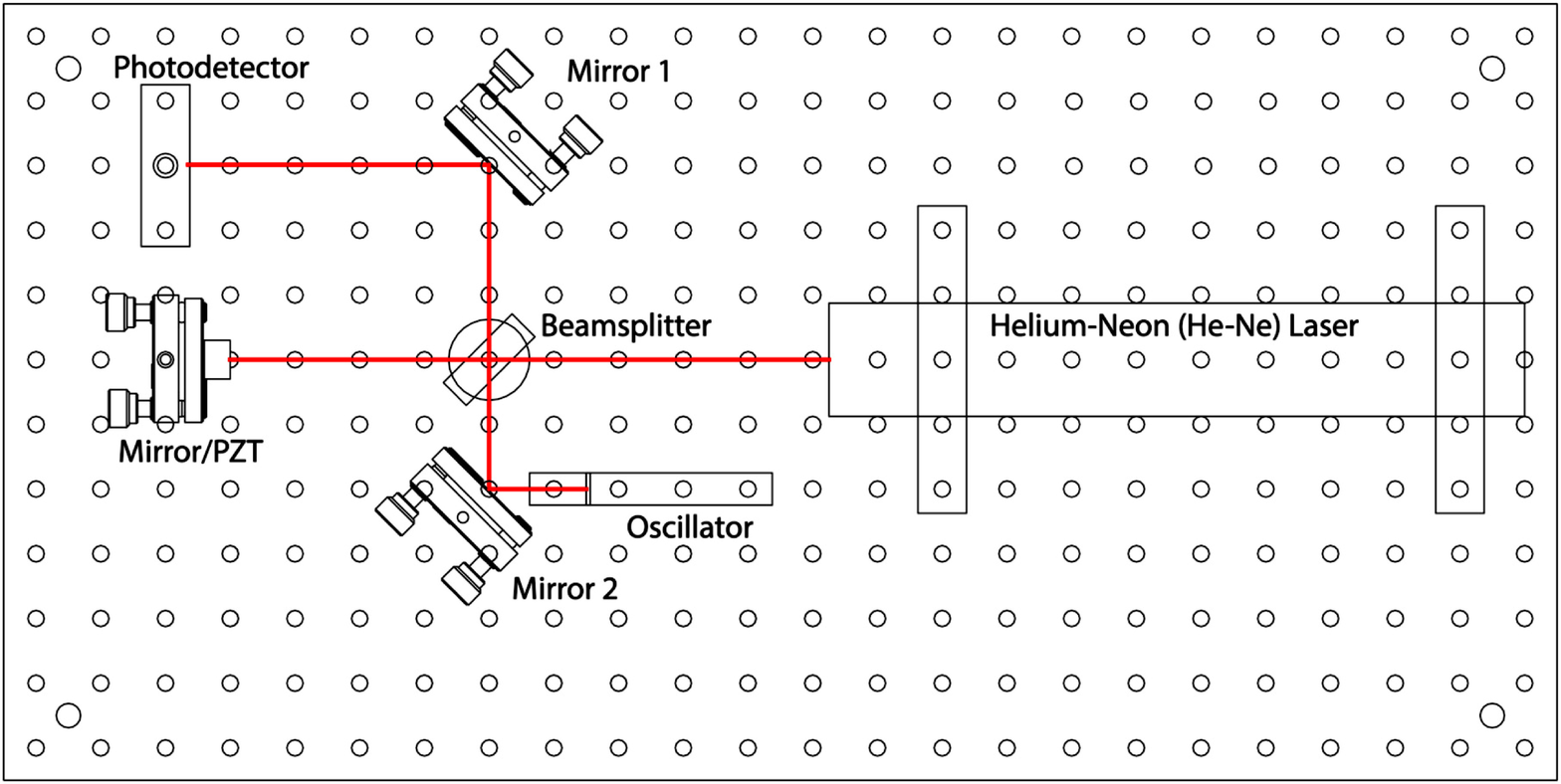}
  \caption{The interferometer optical
layout including the mechanical oscillator shown in detail in Figure \protect
\ref{oscillator}.}
  \label{LayoutOsc}
\end{figure}

\begin{figure}[htb] % float placement: (h)ere, page (t)op, page (b)ottom, other (p)age
  \centering
  % file name: C:/Dropbox/1-kgl-top/Papers/2014/Interferometer/Oscillator.jpg
  \includegraphics[width=2.5in, keepaspectratio]{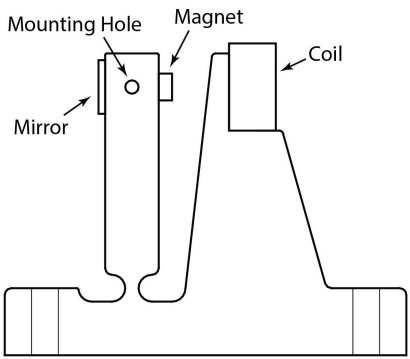}
  \caption{A side view of the magnetically
driven mechanical oscillator shown in Figure \protect\ref{LayoutOsc}. The
main body is constructed from 12.7-mm-thick aluminum plate (alloy 6061), and
the two vertical holes in the base are 76.2 mm apart to match the holes in
the breadboard. Sending an alternating current through the coil applies a
corresponding force to the permanent magnet, driving torsional oscillations
of the mirror arm about its narrow pivot point. Additional weights can be
added to the 8-32 tapped mounting hole to change the resonant frequency of
the oscillator.}
  \label{oscillator}
\end{figure}

Once students have constructed, aligned, and characterized the
interferometer, they can then use it to observe the nanoscale motions of a
simple harmonic oscillator. The optical layout for this second stage of the
experiment is shown in Figure \ref{LayoutOsc}, and the mechanical
construction of the oscillator is shown in Figure \ref{oscillator}. Wiring
for the coil runs through a vertical hole in the aluminum plate (below the
coil but not shown in Figure \ref{oscillator}) and then through one of the
holes in the breadboard to the electronics chassis below. For this reason
the oscillator position on the breadboard cannot be changed, but it does not
interfere with the basic interferometer layout shown in Figure \ref%
{ifolayout}.

\begin{figure}[thb] % float placement: (h)ere, page (t)op, page (b)ottom, other (p)age
  \centering
  % file name: C:/Dropbox/1-kgl-top/Papers/2014/Interferometer/LinearAmp-nm.jpg
  \includegraphics[width=3.5in, keepaspectratio]{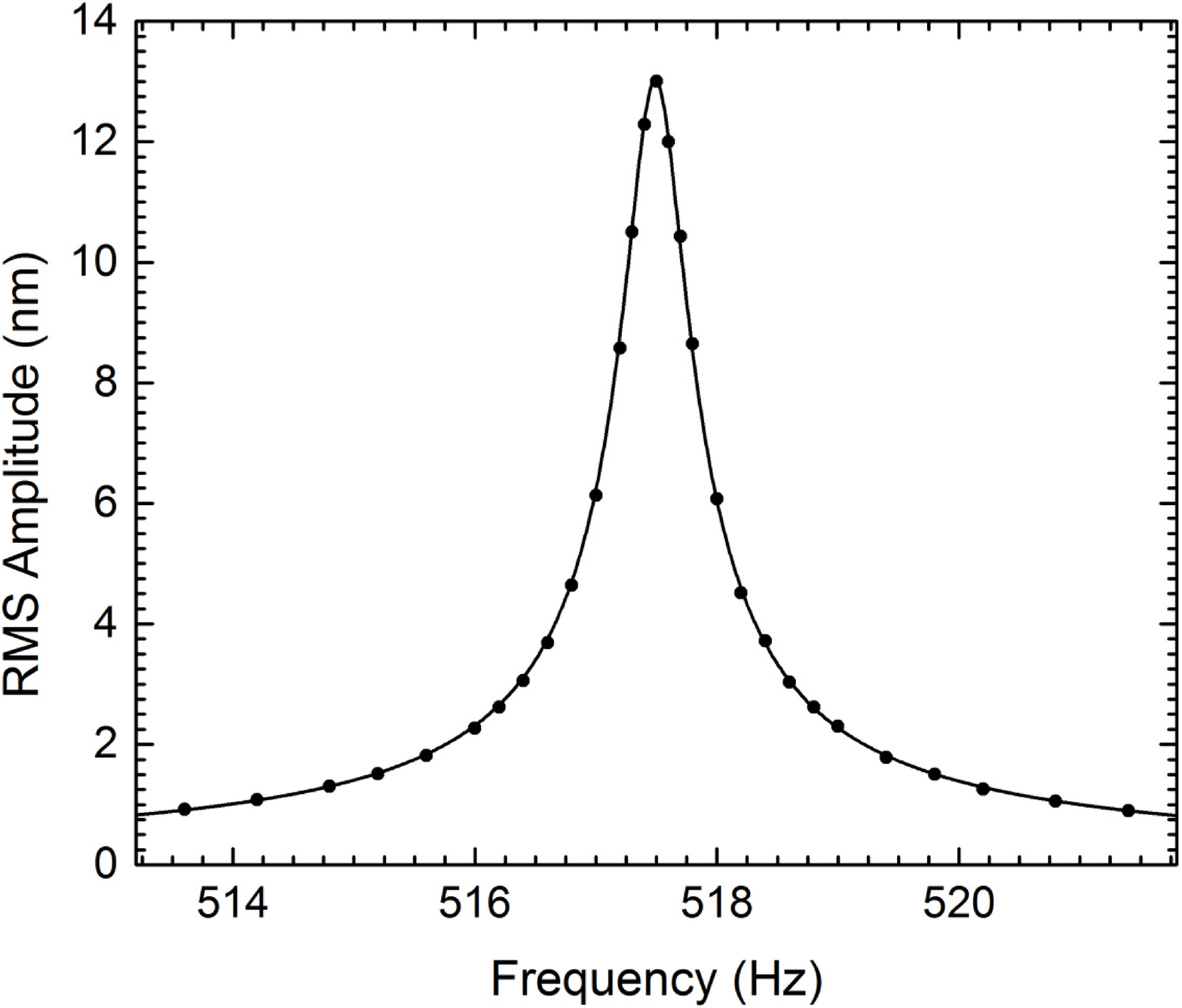}
  \caption{The measured resonant response of
the oscillator as a function of drive frequency. The absolute
root-mean-square (RMS) amplitude was derived optically from the
interferometer signal. The response is well matched by a
simple-harmonic-oscillator model (fit line), indicating a mechanical $Q$ of 970.}
  \label{oscillatorresponse}
\end{figure}

The oscillator response can be observed by viewing the interferometer signal
together with the coil drive signal on the oscilloscope, and example data
are shown in Figure \ref{oscillatorresponse}. Here the coil was driven with
a sinusoidal signal from a digital function generator with $<1$ mHz absolute
frequency accuracy, and the oscillator response was measured for each point
by averaging 64 traces on the oscilloscope. Once again, using the drive
signal to trigger the oscilloscope ensures a good phase-locked average even
with a small signal amplitude. As shown also in Figure \ref{displacement},
sub-nanometer sensitivity is easily achievable using this simple
signal-averaging method. The results in Figure \ref{oscillatorresponse} show
that this mechanical system is well described by a
simple-harmonic-oscillator model. Inserting a small piece of foam between
the magnet and the coil substantially increases the oscillator damping, and
students can examine this by measuring the oscillator $Q$ with different
amounts of damping.

The tapped mounting hole behind the oscillator mirror (see Figure \ref%
{oscillator}) allows additional weights to be added to the oscillator. We
use nylon, aluminium, steel, and brass thumbscrews and nuts to give a series
of weights with roughly equal mass spacings. Students weigh the masses using
an inexpensive digital scale with 0.1 gram accuracy (American Weigh
AWS-100). To achieve satisfactory results, we have found that the weights
need to be well balanced (with one on each side of the oscillator), screwed
in firmly, and no more than about 1.5 cm in total length. If these
conditions are not met, additional mechanical resonances can influence the
oscillator response.

\begin{figure}[thb] % float placement: (h)ere, page (t)op, page (b)ottom, other (p)age
  \centering
  % file name: C:/Dropbox/1-kgl-top/Papers/2014/Interferometer/AddedMass.jpg
  \includegraphics[width=3.5in, keepaspectratio]{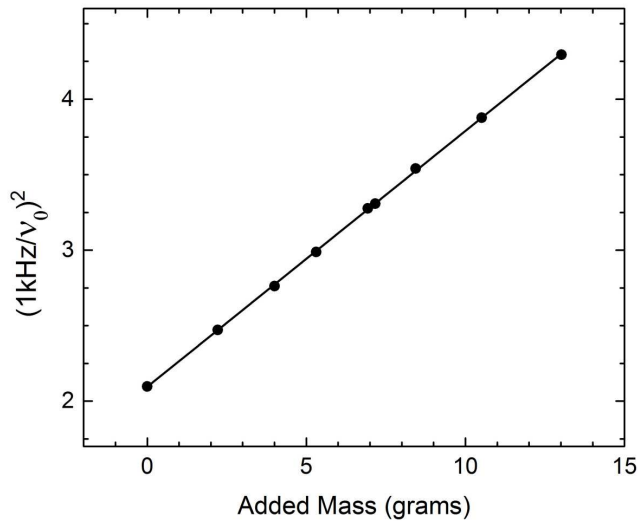}
  \caption{Measured changes in the
resonant frequency $\protect\nu _{0}$ of the oscillator as a function of the
mass added to the mounting hole shown in Figure \protect\ref{oscillator}.
Simple-harmonic-oscillator theory predicts that $\protect\nu _{0}^{-2}$
should scale linearly with added mass. The spring constant and moment of
inertia of the oscillator can be extracted from the slope and intercept of
the fit line.}
  \label{oscillatormass}
\end{figure}

The resonant frequency $\nu _{0}$ of the oscillator can be satisfactorily
measured by finding the maximum oscillator amplitude as a function of
frequency, viewing the signal directly on the oscilloscope, and an accuracy
of better than 1 Hz can be obtained quite quickly with a simple analog
signal generator using the oscilloscope to measure the drive frequency. The
results shown in Figure \ref{oscillatormass} show that $\nu _{0}^{-2}$ is
proportional to the added mass, which is expected from a
simple-harmonic-oscillator model. Additional parameters describing the
harmonic oscillator characteristics can be extracted from the slope and
intercept of the fit line.

As a final experiment, students can drive the coil with a square wave signal
at different frequencies to observe the resulting motion. The oscillator
shows a resonant behavior when the coil is driven at $\nu _{0},$ $\nu
_{0}/3, $ $\nu _{0}/5,$ etc., and at each of these frequencies the
oscillator response remains at $\nu _{0}.$ Measurements of the peak resonant
amplitude at each frequency show the behavior expected from a Fourier
decomposition of the square wave signal.

In summary, we have developed a fairly basic table-top laser interferometer
for use in the undergraduate teaching laboratory. Students first assemble
and align the interferometer, gaining hands-on experience using optical and
laser hardware. The experiment then focuses on a variety of measurement
strategies and signal-averaging techniques, with the goal of using the
interferometer to demonstrate picometer displacement sensitivity over arm
lengths of 10 centimeters. In a second stage of the experiment, students use
the interferometer to quantify the nanoscale motions of a driven harmonic
oscillator system.

This work was supported in part by the California Institute of Technology
and by a generous donation from Dr. Vineer Bhansali. Frank Rice contributed
insightful ideas to several aspects of the interferometer construction and
data analysis.

\end{document}